\documentclass[12pt,floatfix,aps,amsmath,nofootinbib,preprintnumbers,preprint]{revtex4}

\usepackage{graphicx}
\usepackage{bm}

\def\({\left(}
\def\){\right)}
\def\[{\left[}
\def\]{\right]}

\def\e{\begin{equation}}
\def\q{\end{equation}}
\def\m{\begin{eqnarray}}
\def\n{\end{eqnarray}}

\begin{document}

\title{Nearly scale-invariant power spectrum and quantum cosmological perturbations in the gravity's rainbow scenario}

\author{Sai Wang$^{1}$\footnote{E-mail: wangsai@itp.ac.cn}}
\author{Zhe Chang$^{1,2}$}
\affiliation{
${}^1$State Key Laboratory of Theoretical Physics, Institute of Theoretical Physics, Chinese Academy of Sciences, Beijing 100049, China\\
${}^2$Institute of High Energy Physics, Chinese Academy of Sciences, Beijing 100049, China}

\begin{abstract}

We propose the gravity's rainbow scenario as a possible alternative of the inflation paradigm to account for the flatness and horizon problems. We focus on studying the cosmological scalar perturbations which are seeded by the quantum fluctuations in the very early universe. The scalar power spectrum is expected to be nearly scale-invariant. We estimate the rainbow index $\lambda$ and energy scale $M$ in the gravity's rainbow scenario by analyzing the Planck temperature and WMAP polarization datasets. The constraints on them are given by $\lambda=2.933\pm0.012$ and $\ln (10^5M/M_p)= -0.401^{+0.457}_{-0.451}$ at the $68\%$ confidence level.

\end{abstract}


\maketitle


\section{Introduction}

The inflation model \cite{Guth:1980zm,Linde:1981mu,Albrecht:1982wi} has been the leading paradigm for the very early universe in the last three decades. In the inflation paradigm, the scale factor $a(t)$ of the universe undergone a stage of exponential expansion in a very short time. This leads the universe to be flat enough to account for the flatness problem, since $|\Omega_K|\propto a^{-2}$. On the other hand, this reveals that the cosmological scales observed today were deep inside the Hubble scale, which accounts for the horizon problem. Moreover, the cosmological scalar perturbations can be seeded by the primordial quantum fluctuations which are stretched outside of the horizon (see \cite{Mukhanov:1990me} for reviews). The scalar power spectrum is predicted to be adiabatic, Gaussian, and nearly scale-invariant. This is well consistent with the present astronomical observations on the anisotropy of cosmic microwave background (CMB) and the formation of large-scale structures (LSS). Although the inflation fits the observational data well, it still suffers several significant issues, such as the fine-tuning slow-roll potential \cite{Penrose1989}, the initial conditions \cite{Borde:2001nh,Carroll:2005it}, and the trans-Planckian problem \cite{Brandenberger:2012aj}, etc.. In addition, one requires an inflaton field to drive the exponential expansion of the very early universe. However, the astronomical observations have not yet discovered such a fundamental scalar field until recently.

It is interesting to study possible alternatives for the inflation paradigm. In an alternative scenario, actually, one just need require that the observed universe were inside the particle horizon in the very early universe to account for the problems of big-bang cosmology. In this paper, we propose that the gravity's rainbow scenario shall meet this requirement. The gravity's rainbow scenario \cite{Magueijo:2002xx} is arisen from the phenomenological studies of the quantum gravity which should play a significant role in the very early universe. Recently, it has been utilized to study the very early universe \cite{Weinfurtner:2008if,Ling:2008sy,Corda:2010uq,Garattini:2012ca,Majumder:2013mza,Amelino-Camelia:2013wha,Amelino-Camelia:2013tla,Mukohyama:2009gg,Magueijo:2008yk,Awad:2013nxa,Barrow:2013gia}. The spacetime metric felt by a free particle would be dependent on the energy (or momentum equivalently) of the particle in the gravity's rainbow scenario. Thus the dispersion relation can be significantly modified for a ultra-relativistic particle. This leads to an effective speed of light. The varying speed of light (VSL) cosmology has been proposed \cite{Moffat:1992ud,Albrecht:1998ir,Barrow:1999jq}, and the observable universe was assumed to be only a part of the causal area if the effective speed of light is large enough in the very early universe. Thus, the gravity's rainbow scenario shall be potential to resolve the flatness and horizon problems.

In the gravity's rainbow scenario, the evolution of the very early universe would be driven by the thermally fluid substance instead of a fundamental scalar field. We will study the thermodynamics of the system of ultra-relativistic particles with the modified dispersion relation. Then the background evolution of the universe is determined by the modified Friedmann equation. The solution of the Friedmann equation will be showed to resolve the flatness and horizon problems. We shall focus on studying the cosmological linear perturbations and their quantization in this paper. The issue of gauge choices will be studied in detail, and then the perturbed Einstein's field equations will be calculated in the longitudinal gauge. We will construct the comoving curvature perturbation which is gauge-invariant and conserved outside the Hubble horizon. In this model, the quantum fluctuations are expected to be dominated above the rainbow energy scale.
Furthermore, we will make constraints on the parameters of the gravity's rainbow effects by a joint analysis of the Planck temperature \cite{Ade:2013zuv} and WMAP polarization \cite{Hinshaw:2012aka} datasets.


The rest of the paper is arranged as follows. In section II, we study the evolution of the background spacetime in the gravity's rainbow scenario. In section III, the equations of motion are derived for the scalar perturbations. We quantize the scalar perturbations in the longitudinal gauge in section IV. In section V, we calculate the power spectrum of the primordial scalar perturbations and then make constraints on the parameters of the gravity's rainbow effects. The conclusions and discussions are given in section VI.

\section{Evolution of background}

The gravity's rainbow scenario was originally studied by Magueijo \& Smolin \cite{Magueijo:2002xx}. The spacetime metric felt by a free particle depends on the energy or momentum of the particle. In the study of cosmology, we are interested in the spatially flat Friedmann-Robertson-Walker (FRW) metric which is homogeneous and isotropic. In this paper, we will study the evolution of the very early universe with the modified Friedmann-Robertson-Walker (FRW) metric of the form \cite{Ling:2006az}
\begin{equation}
\label{rainbow metric}
d\tau^2=c^2(p)dt^2-a^2(t)\delta_{ij}dx^i dx^j\ ,
\end{equation}
where $a(t)$ is the scale factor of the universe,
the rainbow function $c(p)$ is explicitly parameterized as a power-law form, namely,
\begin{equation}
\label{rainbow function}
c(p)=1+\left({p}/{M}\right)^\lambda\ .
\end{equation}
Here $M$ is an energy scale related to the quantum gravity, and $\lambda$ is called the rainbow index which is positive. The rainbow function $c(p)$ takes the limit $\lim_{p/M\rightarrow 0}c(p)=1$.
In the tangent space, the metric (\ref{rainbow metric}) would lead to the modified dispersion relation for a free ultra-relativistic particle. This could be given as
\begin{equation}
\label{MDR}
E = c(p)p=p+M^{-\lambda}p^{\lambda+1}\ ,
\end{equation}
where we have neglected the mass term for the particle, since the particle's mass is tiny compared to the ultra-relativistic energy.

In the enough early era of the universe, the particle could have an extremely high energy scale, i.e., $p\gg M$.
Then the rainbow function becomes
\begin{equation}
\label{rainbow function p}
c(p)\simeq \left(p/M\right)^{\lambda}\ .
\end{equation}
Thus the second term at the right hand side in (\ref{MDR}) will be dominated, namely,
\begin{equation}
\label{MDRp}
E\simeq M^{-\lambda}p^{\lambda+1}\ .
\end{equation}
Consider a system of such ultra-relativistic particles in thermal equilibrium. It should meet the Maxwell-Boltzmann distribution. We can obtain the energy density $\rho(T)$ of the system with the temperature $T$, namely,
\begin{eqnarray}
\label{energy density}
\rho(T)&=&\int_{0}^{\infty} 4\pi p^2 E e^{-E/T} dp\ \nonumber\\
&=&\frac{4\pi}{\lambda+1}M^{\frac{3\lambda}{\lambda+1}}\int_{0}^{\infty}
E^{\frac{3}{\lambda+1}}e^{-E/T}dE\ \nonumber\\
&=& \sigma(\lambda,M) T^{\frac{\lambda+4}{\lambda+1}} \propto T^{\frac{\lambda+4}{\lambda+1}}\ ,
\end{eqnarray}
where the constant coefficient $\sigma(\lambda,M)$ is given as
\begin{equation}
\sigma(\lambda,M)=\frac{4\pi}{\lambda+1}\Gamma(\frac{\lambda+4}{\lambda+1})
M^{\frac{3\lambda}{\lambda+1}}\ .
\end{equation}
In the second equality of (\ref{energy density}), we used the relation (\ref{MDRp}).
The pressure $P(T)$ of the system could be obtained by resolving the ordinary differential equation
\begin{equation}
T\frac{dP}{dT}-P=\rho\ .
\end{equation}
Thus, it is given by
\begin{equation}
\label{pressure}
P(T)=\frac{\lambda+1}{3}\sigma T^{\frac{\lambda+4}{\lambda+1}}=\omega\rho(T)\ ,
\end{equation}
where we neglected an integral constant, and the state parameter is given by
\begin{equation}
\omega=\frac{\lambda+1}{3}\ .
\end{equation}
The relation (\ref{pressure}) is just the so-called the equation of state.
In addition, the speed of sound could be obtained as $c_s^2:=\frac{\partial P}{\partial \rho}=\frac{\lambda+1}{3}$ which is also a constant. When $\lambda=0$, the above results on $\rho$, $P$, $w$ and $c_s$ would return back to the conventional form for the massless particles (such as the photons) in special relativity.

The conservation of energy-momentum tensor gives the equation of continuity for the thermodynamic system. For a system of ideal fluid, the energy-momentum tensor is given by
\begin{equation}
\label{energy momentum tensor}
T^{\mu}_{\nu}=\left(\rho+P\right)u^{\mu}u_{\nu}-P\delta^{\mu}_{\nu}\ ,
\end{equation}
where $u^{\mu}u_{\mu}=1$. Its conservation implies the equation $T^{\mu}_{0;\mu}=0$. Implicitly, this equation can be written as the equation of continuity
\begin{equation}
\label{continuity}
\frac{d\rho}{dt}+3H\left(\rho+P\right)=0\ ,
\end{equation}
where the Hubble parameter $H=\dot{a}/a$ and $\dot{a}={da}/{dt}$. By combining (\ref{pressure}) with (\ref{continuity}), we obtain
\begin{equation}
\label{rho a}
\rho=\rho_e \left(\frac{a}{a_e}\right)^{-3(1+\omega)}\propto a^{-(\lambda+4)}\ .
\end{equation}
Hereafter the subscript ``e'' denotes physical quantities at the moment when the gravity's rainbow effect is no longer dominated, i.e., $p_e\simeq M$. By comparing (\ref{rho a}) to (\ref{energy density}), we obtain a useful relation, i.e.,
\begin{equation}
\label{a T}
a \propto T^{-\frac{1}{\lambda+1}}\ ,
\end{equation}
or equivalently, $a=a_e\left(T/T_e\right)^{-\frac{1}{\lambda+1}}
= a_e\left(T/M\right)^{-\frac{1}{\lambda+1}}$. Here $a_e$ can be roughly estimated at the temperature $T_e=M$ by the $\Lambda$CDM model.

The evolution of the scale factor $a(t)$ is determined by the Friedmann equation which is deduced from the Einstein's field equation \cite{Mukhanov:2005sc}. In this paper, we assume the modified Einstein's equation as follows
\begin{equation}
\label{Einstein's equation}
G^{\mu}_{\nu}=\frac{1}{c^4} T^{\mu}_{\nu}\ ,
\end{equation}
where $G_{\mu\nu}$ is the Einstein tensor. In this paper, we set $M_p^{-2}=8\pi G=1$.
The rainbow function $c$ appears as the effective speed of light in the above equation.
Then the 00-component of Einstein's equation gives the Friedmann equation
\begin{equation}
\label{Friedmann equation}
H^{2}=\frac{1}{3c^2}\rho\ .
\end{equation}
In the Friedmann equation, one takes the ultra-relativistic particles as an ensemble rather than picking out a specific particle randomly \cite{Ling:2006az}. One should take into account the average effects of the ensemble on the evolution of the very early universe. By using (\ref{rainbow function p}) and (\ref{MDRp}), we obtain $c(E)=\left(E/M\right)^{\frac{\lambda}{\lambda+1}}$.
As an ensemble, the thermodynamic system in thermal equilibrium has a typical energy scale, namely, the temperature $T$ takes a statistical mean value.
Thus, one could take $T$ as the energy appearing in the gravity's rainbow metric,
namely, 
\begin{equation}
\label{c(T)}
c\equiv c(T)=\left(T/M\right)^{\frac{\lambda}{\lambda+1}}\ .
\end{equation}
Then we could resolve the Friedmann equation (\ref{Friedmann equation}), and the solution is
\begin{equation}
\label{a t}
a(t)\propto t^{\frac{2}{4-\lambda}}\ ,
\end{equation}
where we used (\ref{a T}) and (\ref{c(T)}).
We set $\lambda<4$ to obtain an expanding universe, while $\lambda>4$ is related to
a contracting universe.
If $\lambda=4$, the exponent $2/(4-\lambda)$ would be divergent.
This case is not well-defined.

The flatness and horizon problems can be demonstrated as follows. The spatial curvature term $|\Omega_k|=\frac{c^2}{a^2H^2}$ is proportional to $T^{\frac{3\lambda-2}{\lambda+1}}$. With the decrease of temperature in the expanding universe, $|\Omega_k|$ should also decrease across more than $24$ orders of magnitude to resolve the flatness problem. Then we require $\lambda>2/3$ and a high energy scale $T_i\gg T_e$. Hereafter the subscript ``i'' denotes the start time of the rainbow universe. On the other hand, the particle horizon $d_H=\int_{t_i}^{t_e}\frac{c dt}{a}=\int^{a_e}_{a_i}\frac{cda}{a^2 H}$ is proportional to $T^{\frac{3\lambda-2}{2(\lambda+1)}}$. Similarly to resolving the flatness issue, it also requires $\lambda>2/3$ to resolve the horizon problem. If $\lambda>4$, however, the temperature would increase with increase of the time based on ($\ref{a T}$) and (\ref{a t}). This makes even worse the flatness and horizon problems. Thus, the above discussions show that the rainbow index $\lambda$ should satisfy the condition $2/3<\lambda<4$.

\section{Scalar perturbations}

In the following, we shall focus on studying the cosmological linear perturbations and their quantization while disregarding the statistically thermal fluctuations \cite{footnote}. The rainbow function $c(T)$ could be formally viewed as a smooth background function of the temperature. Consider the scalar perturbations. The perturbed rainbow metric takes the form
\begin{eqnarray}
\label{scalar perturbations general}
d\tau^2&=&a^2[(1+2\phi)c^2d\eta^2+2cB_{,i}dx^i d\eta-\nonumber\\
&~&~~~~\left((1-2\psi)\delta_{ij}-2E_{,ij}\right) dx^i dx^j] \ ,
\end{eqnarray}
where we have used the conformal time $d\eta=dt/a$. We consider the coordinate transformation $x^\mu\rightarrow \tilde{x}^\mu=x^\mu+\zeta^\mu$. Here $\zeta^\mu=(\zeta^0,\zeta^i)$ denote the infinitesimal functions of the spacetime coordinates, and $\zeta^i=\zeta^i_\perp +\varsigma^{,i}$ where $\zeta^i_{\perp,i}=0$ and $\varsigma$ denotes a scalar function. The Lie derivative of the metric perturbations $\delta g_{\mu\nu}(x)=g_{\mu\nu}(x)-\bar{g}_{\mu\nu}(x)$ gives the gauge transformation law, i.e., $\delta g_{\mu\nu}\rightarrow \delta\tilde{g}_{\mu\nu}=\delta g_{\mu\nu}-\bar{g}_{\mu\nu,\sigma}\zeta^\sigma-\bar{g}_{\sigma\nu}\zeta^\sigma_{,\mu}-\bar{g}_{\mu\sigma}\zeta^\sigma_{,\nu}$, where $\bar{g}_{\mu\nu}$ denotes the unperturbed background metric. Thus, we can obtain
\begin{eqnarray}
\label{scalar perturbation gauge transform}
&&\phi\rightarrow\tilde{\phi}=\phi-\frac{1}{ac}\left(ac\zeta^0\right)^\prime\ ,\\
&&\psi\rightarrow\tilde{\psi}=\psi+\frac{a^\prime}{a}\zeta^0\ ,\\
&&B\rightarrow \tilde{B}=B+\frac{1}{c}\varsigma^\prime-c\zeta^0\ ,\\
&&E\rightarrow\tilde{E}=E+\varsigma\ .
\end{eqnarray}
Hereafter, the primes denotes the derivative with respect to the conformal time $\eta$.
The Bardeen's potentials are the simplest gauge-invariant linear combinations of the above scalar perturbations. They are given by
\begin{eqnarray}
\label{Bardeen's potentials}
&&\Phi=\phi-\frac{1}{ac}\left[a\left(B-\frac{1}{c}E^\prime\right)\right]^\prime\ ,\\
&&\Psi=\psi+\frac{a^\prime}{ac}\left(B-\frac{1}{c}E^\prime\right)\ .
\end{eqnarray}
In the longitudinal gauge, we choose the system of coordinates with $B=E=0$. Thus, the perturbed rainbow metric (\ref{scalar perturbations general}) can be rewritten as
\begin{equation}
\label{rainbow metric in longitudinal gauge}
d\tau^2=a^2\left[(1+2\Phi)c^2 d\eta^2-(1-2\Psi)\delta_{ij}dx^i dx^j\right]\ .
\end{equation}
If the anisotropic stresses are not considered, we can obtain the relation $\Psi=\Phi$ as will be demonstrated later.

To derive the equations for the linear cosmological perturbations, we can linearize the Einstein's field equations
\begin{equation}
\label{perturbed Einstein equations}
\delta G^{\mu}_{\nu}=\frac{1}{c^4}\delta T^{\mu}_{\nu} \ ,
\end{equation}
where $\delta G^{\mu}_{\nu}$ and $\delta T^{\mu}_{\nu}$ denote the gauge-invariant perturbations. In general, the perturbed energy-momentum tensor is given by
\begin{equation}
\label{perturbed energy-momentum tensor}
\delta T^0_0= \delta \rho~,~~\delta T^0_i=\rho (1+\omega) v_{,i}~,~~\delta T^i_j=- c_s^2 \delta \rho\delta^i_j~\ ,
\end{equation}
where we have neglected the anisotropic stress. Only the adiabatic perturbations are considered in this paper. Thus, the spatial components of the Einstein's field equations can be explicitly given by
\begin{eqnarray}
\left[\Psi^{\prime\prime}+\mathcal{H}(2\Psi+\Phi)^\prime+(2\mathcal{H}^\prime+\mathcal{H}^2)\Phi+\frac{c^2}{2}\Delta(\Phi-\Psi)\right]\delta^i_j~~&&\nonumber\\
-\mathcal{H}\frac{d\ln c}{d\ln a}\left(\Psi^\prime+2\mathcal{H}\Phi\right)\delta^i_j
-\frac{c^2}{2}(\Phi-\Psi)^{,i}_{,j}
=-\frac{1}{2c^2} a^2 \delta T^i_j\ .~&&
\label{e33}
\end{eqnarray}
Here $\mathcal{H}=a^\prime/a$ denotes the comoving Hubble parameter.
For $i\neq j$, we have $\delta T^i_j=0$, and then (\ref{e33}) is reduced to $(\Phi-\Psi)_{,ij}=0$. The only solution is $\Psi=\Phi$, which is similar to the result in the standard model \cite{Mukhanov:2005sc}. By considering this result, we obtain the following equations for the scalar perturbations
\begin{eqnarray}
\label{e1}
c^2 \Delta\Phi-3\mathcal{H}(\Phi^\prime+\mathcal{H}\Phi)=\frac{1}{2c^2} a^2 \delta \rho\ ,~~~&&\\
\label{e2}
(\Phi^\prime+\mathcal{H}\Phi)_{,i}=\frac{1}{2c^3} a^2 \rho (1+\omega) v_{,i}\ ,~~~&&\\
\label{e3}
\left[\Phi^{\prime\prime}+3\mathcal{H}\Phi^\prime+(2\mathcal{H}^\prime+\mathcal{H}^2)\Phi\right]
-\mathcal{H}\frac{d\ln c}{d\ln a}\left(\Phi^\prime+2\mathcal{H}\Phi\right)~~~&&\nonumber\\
=\frac{1}{2c^2} a^2 c_s^2 \delta \rho\ .~~~&&
\end{eqnarray}
By combining (\ref{e1}) and (\ref{e3}), we obtain an equation for the gravitational potential $\Phi$, namely,
\begin{eqnarray}
\label{phi equation}
&&\Phi^{\prime\prime}+3\mathcal{H}\left(1+c_s^2-\frac{1}{3}\frac{d\ln c}{d\ln a}\right)\Phi^\prime - c_s^2 c^2 \Delta \Phi\nonumber\\
&&+\left[2\mathcal{H}^\prime+\mathcal{H}^2\left(1+3c_s^2-2\frac{d\ln c}{d\ln a}\right)\right]\Phi=0\ .
\end{eqnarray}

Before resolving (\ref{phi equation}), we shall discuss the comoving curvature perturbation. This is a gauge-invariant quantity which is conserved outside the Hubble horizon.
In general, it is defined by
\begin{equation}
\label{curvature perturbation}
\mathcal{R}=-\Phi-\frac{1}{c}\mathcal{H}v\ .
\end{equation}
Outside the Hubble horizon, one can disregard the terms proportional to $\Delta \Phi$. By combining (\ref{e1}) and (\ref{e2}), thus, one gets the equation $c\delta \rho+3\mathcal{H}(\rho+P)v=0$. Therefore, the comoving curvature perturbation can be rewritten as
\begin{equation}
\label{comoving curvature perturbation}
\mathcal{R}=-\Phi-\frac{\delta \rho}{3(\rho+P)}\ .
\end{equation}
Its derivative with respect to time is given by
\begin{equation}
\label{Rp}
\mathcal{R}^\prime=\frac{\rho^\prime\delta P-P^\prime\delta\rho}{3(\rho+P)^2}\ ,
\end{equation}
where we have used the equations for the energy-momentum conservation, i.e., $\rho^\prime+3\mathcal{H}(\rho+P)=0$ and $\delta\rho^\prime+3\mathcal{H}(\delta\rho+\delta P)-3(\rho+P)\Phi^\prime=0$. Noting $P=\omega \rho$ and $\omega$ is a constant, the right-handed term must vanish in (\ref{Rp}). Thus, $\mathcal{R}$ is conserved outside the Hubble horizon.

\section{Quantizing perturbations}

The equation (\ref{phi equation}) for the gravitational potential $\Phi$ can be reduced into a simpler form. In this paper, we just consider the case of $\lambda>2$, for which the reason will be clear later. By noting $a\propto (-\eta)^{\frac{2}{2-\lambda}}$, $c_s^2=\frac{1+\lambda}{3}$ and $c\propto a^{-\lambda}$, we obtain the equation
\begin{equation}
\label{phi equation eta}
\Phi^{\prime\prime}+{2q}{\eta^{-1}}\Phi^\prime+\left(-{m}{\eta^{-\ell}}\Delta+{\bar{n}}{\eta^{-2}}\right)\Phi=0\ ,
\end{equation}
where $q=\frac{2(2+\lambda)}{2-\lambda}$, $\ell=\frac{4\lambda}{2-\lambda}$, $\bar{m}=\frac{\lambda+1}{3} \eta_e^\ell$ and $\bar{n}=\frac{16\lambda}{(2-\lambda)^2}$.
Here we have chosen the end moment of the gravity's rainbow effects as the original point of time. Hereafter, the subscript ``e'' denotes the quantity at the moment when the gravity's rainbow effects become to be no longer dominated. One can introduce a new variable to eliminate the term proportional to $\Phi^\prime$. The new variable is given by $u=(-\eta/\eta_e)^q \Phi$. Then the equation (\ref{phi equation eta}) becomes
\begin{equation}
\label{uk}
u^{\prime\prime}-\left({m}{\bar{\eta}^{-\ell}}\Delta+{n}{\bar{\eta}^{-2}}\right)u=0\ ,
\end{equation}
where $\bar{\eta}=-\eta/\eta_e$, $m=\frac{\lambda+1}{3}\eta_e^2$, and $n=q(q-1)-\bar{n}={2(4+3\lambda^2)}/{(2-\lambda)^2}$. Hereafter the prime denotes the derivatives with respect to $\bar{\eta}$.
To quantize the new perturbation $u$, the equation (\ref{uk}) can be corresponded to the action of the form
\begin{equation}
\label{action}
S=\int \mathcal{L}d\bar{\eta} dx^3=\int \frac{1}{2}\left(u^{\prime 2}+m\bar{\eta}^{-\ell}u\Delta u+n\bar{\eta}^{-2}u^2\right) d{\bar{\eta}} dx^3 \ ,
\end{equation}
which is very different from the one in the inflation paradigm. The canonical momentum conjugated to $u$ is defined as $\pi=\partial \mathcal{L}/\partial u^\prime= u^\prime$.

In the quantization process, the field variable $u$ and its canonical momentum $\pi$ become operators $\hat{u}$ and $\hat{\pi}$, respectively.
The operator $\hat{u}$ obeys the equation
\begin{equation}
\label{uhat}
\hat{u}^{\prime\prime}-\left({m}{\bar{\eta}^{-\ell}}\Delta+{n}{\bar{\eta}^{-2}}\right)\hat{u}=0\ ,
\end{equation}
which is same as (\ref{uk}). In general, the solution of the above equation can be given by
\begin{equation}
\label{uk solution}
\hat{u}(\bar{\eta},\mathbf{x})=\int\frac{d^3\mathbf{k}}{(2\pi)^{3/2}} \frac{1}{\sqrt{2}}\left(u^\ast_\mathbf{k}(\bar{\eta})e^{i\mathbf{kx}}\hat{a}_\mathbf{k}^{-}+u_\mathbf{k}(\bar{\eta})e^{-i\mathbf{kx}}\hat{a}_\mathbf{k}^{+}\right)\ ,
\end{equation}
where $u_\mathbf{k}(\bar{\eta})$ satisfies
\begin{equation}
\label{uk1}
u_\mathbf{k}^{\prime\prime}+\left({m}{\bar{\eta}^{-\ell}}\mathbf{k}^2-{n}{\bar{\eta}^{-2}}\right)u_\mathbf{k}=0\ ,
\end{equation}
and the bosonic commutation relations are given for the creation and annihilation operators as follows
\begin{eqnarray}
\label{a}
[\hat{a}^{-}_\mathbf{k},\hat{a}^{-}_\mathbf{k^\prime}]=[\hat{a}^{+}_\mathbf{k},\hat{a}^{+}_\mathbf{k^\prime}]=0,~~[\hat{a}^-_\mathbf{k},\hat{a}^{+}_\mathbf{k^\prime}]=\delta^{(3)}(\mathbf{k}-\mathbf{k^\prime})\ .
\end{eqnarray}
The vacuum $|0\rangle$ is defined as the state which is annihilated by $a_{\mathbf{k}}^{-}$, i.e., $a_{\mathbf{k}}^{-}|0\rangle=0$.
One requires $u_\mathbf{k}(\eta)$ to satisfy the normalization condition
\begin{equation}
\label{uk normalization condition}
u^\prime_{\mathbf{k}}u^\ast_{\mathbf{k}}-u_{\mathbf{k}}u^{\ast\prime}_{\mathbf{k}}=2 i\ ,
\end{equation}
of which the left-hand-side term is the Wronskian of (\ref{uk}).
At any a given time, thus, the operators $\hat{u}$ and $\hat{\pi}$ satisfy the commutation relations, i.e.,
\begin{eqnarray}
\label{u commutation}
&&[\hat{u}(\bar{\eta},\mathbf{x}),\hat{u}(\bar{\eta},\mathbf{y})]=[\hat{\pi}(\bar{\eta},\mathbf{x}),\hat{\pi}(\bar{\eta},\mathbf{y})]=0\ ,\\
&&[\hat{u}(\bar{\eta},\mathbf{x}),\hat{\pi}(\bar{\eta},\mathbf{y})]=i\delta^{(3)}(\mathbf{x}-\mathbf{y})\ .
\end{eqnarray}
The equation (\ref{uk}) has two independent solutions which are represented in terms of the Bessel functions, namely,
\begin{eqnarray}
u_\mathbf{k}^{(1)}(\bar{\eta})&=&\bar{\eta}^{\frac{1}{2}}J_{\frac{\sqrt{1+4n}}{2-\ell}}(\frac{2\sqrt{m}{k}}{\ell-2}\bar{\eta}^{\frac{2-\ell}{2}})\ ,\\
u_\mathbf{k}^{(2)}(\bar{\eta})&=&\bar{\eta}^{\frac{1}{2}}Y_{\frac{\sqrt{1+4n}}{2-\ell}}(\frac{2\sqrt{m}{k}}{\ell-2}\bar{\eta}^{\frac{2-\ell}{2}}) \ ,
\end{eqnarray}
where we denote $k^2=\mathbf{k}^2$.
Thus, its general solution can be expressed as
\begin{equation}
\label{general solution}
u_\mathbf{k}(\bar{\eta})=c_1 u_\mathbf{k}^{(1)}(\bar{\eta})+c_2 u_\mathbf{k}^{(2)}(\bar{\eta})\ .
\end{equation}
Note the Abel's identity $J_\alpha(x)\frac{dY_\alpha(x)}{dx}-\frac{dJ_\alpha(x)}{dx}Y_\alpha(x)=\frac{2}{\pi x}$ for the Bessel functions. Therefore, we could formally give the coefficients $c_1$ and $c_2$ as follows
\begin{equation}
\label{c1c2}
c_1=\sqrt{\frac{\pi\eta_e}{2-\ell}}\ ,~~c_2=-ic_1\ .
\end{equation}
Here we have disregarded a common complex-number factor which is unitary.
In the UV regime, if $\lambda=0$, the above solution coincides with the standard formula $u_\mathbf{k}\sim\frac{1}{\sqrt{c_sk}}e^{-ic_sk\eta}$ in the Minkowski spacetime.
The reason is that $J_\alpha(x)-iY_\alpha(x)$ has the asymptotic expression which is proportional to $\sqrt{\frac{2}{\pi x}}e^{-ix}$ when $x>>0$.
By substituting (\ref{c1c2}) into (\ref{general solution}), we obtain the general representation for $u_\mathbf{k}(\bar{\eta})$.

\section{Primordial power spectrum}

We are particularly interested in the long-wavelength perturbations. At the initial moment, these modes are deeply inside the Hubble horizon because of the large quantity for the effective speed of light. With decrease of the temperature, the effective speed of light decrease rapidly. Thus, these modes would exit from the Hubble horizon. After the dominating era of the gravity's rainbow effects, they reentered the Hubble horizon with the expansion of the universe. In the IR regime, the Bessel function with $\alpha>0$ has the asymptotic expression $J_\alpha(x)\rightarrow\frac{1}{\Gamma(\alpha+1)}\left(\frac{x}{2}\right)^\alpha$ and $Y_\alpha(x)\rightarrow-\frac{\Gamma(\alpha)}{\pi}\left(\frac{2}{x}\right)^\alpha$. Thus, the term in $Y_\alpha(x)$ will be dominated for the long-wavelength perturbations. Therefore, the power spectrum for the gravitational potential $\Phi$ is given by
\begin{eqnarray}
\label{PW for GP}
P_\Phi(k)&=&\frac{k^3}{2\pi^2}\langle \Phi_\mathbf{k}^\ast \Phi_\mathbf{k}\rangle\propto{k^3}\eta^{-2q}\langle u_\mathbf{k}^\ast u_\mathbf{k}\rangle\nonumber\\
&\propto&k^{3-2\frac{\sqrt{1+4n}}{\ell-2}}\eta^{1-2q-\sqrt{1+4n}}\nonumber\\
&\propto&k^{\frac{4(\lambda-3)}{3\lambda-2}}\ ,
\end{eqnarray}
where we have used $u\propto (-\eta)^q \Phi$, and in the last step the relation $k=\mathcal{H}/c\propto a^{\frac{3\lambda-2}{2}}\propto (-\eta)^{\frac{3\lambda-2}{2-\lambda}}$ for the horizon-crossing modes. A more detailed calculation can give the amplitude for the above power spectrum of gravitational potential. In fact, the amplitude  $A_\Phi$ is given by $\tiny{\frac{3\Gamma^2(\alpha)}{2\pi^3(\lambda+1)}\left(\frac{\lambda-2}{2}\right)^{\frac{-2(\lambda+4)}{3\lambda-2}}k_{\rm{pivot}}^{\frac{4(\lambda-3)}{3\lambda-2}}M^{\frac{2(\lambda+4)}{3\lambda-2}}\left(\frac{3(3\lambda-2)}{\lambda+1}\right)^{2\alpha-1}}$, where $\alpha=\frac{\sqrt{25\lambda^2-4\lambda+36}}{2(3\lambda-2)}$ and $k_{\rm{pivot}}$ denotes a pivot scale.
The amplitude $A_\Phi$ is proportional to $M^2$ for the scale-invariant spectrum.
To roughly estimate the magnitude of quantities for the energy scale $M$, one could approximate the amplitude as $A_\Phi\simeq M^{\frac{2(\lambda+4)}{3\lambda-2}}$.
Under the limit $\lambda=0$, the universe becomes radiation-dominated in our model. The cosmological perturbations will be mainly determined by the fluctuations of radiations, whose amplitude $\zeta$ is scaled as $1/\sqrt{V}$. Here $V$ denotes a volume related to Hubble horizon. On the other hand, the volume $V$ is proportional to $k^{-3}$ for the horizon-crossing modes. Here $k$ denotes the wavenumber of a horizon-crossing mode. As a convention, the power spectrum of cosmological perturbations is usually defined as $P(k)=\frac{k^3}{2\pi^2}|\zeta|^2$ in cosmology. Therefore, we can obtain the power spectrum which is given by $P(k)\sim k^3 (1/\sqrt{V})^2\sim k^6$ under the limit $\lambda=0$.

Once the power spectrum for the gravitational potential is got, we can immediately obtain the power spectrum for the comoving curvature perturbation. By using (\ref{e2}) and (\ref{curvature perturbation}) and transforming to the momentum space, we can represent the comoving curvature perturbation (\ref{comoving curvature perturbation}) in terms of the gravitational potential $\Phi_\mathbf{k}$, namely,
\begin{equation}
\label{curvature perturbation gravity}
\mathcal{R}_\mathbf{k}=-\Phi_\mathbf{k}\left[1+\frac{2}{3(1+\omega)}\left(1+\frac{\Phi_\mathbf{k}^\prime/\Phi_\mathbf{k}}{a^\prime/a}\right)\right] \ ,
\end{equation}
where we have used the Friedmann equation. The term in the square bracket is a constant in the above representation. We denote it by $A$. Thus, the power spectrum of $\mathcal{R}_\mathbf{k}$ could be given by $P_{\mathcal{R}}(k)=\left(A^{\ast} A\right) P_{\Phi}(k)$.
Here we can calculate $A$ by using $\Phi\propto\eta^{-q}u$ and the asymptotic IR expression of $u_\mathbf{k}(\eta)$, and the result is
$A=\frac{1}{2(\lambda+4)}\left[\sqrt{25\lambda^2-4\lambda+36}-3(\lambda-2)\right]$.
Once $\lambda$ is determined, one can calculate $A$ and then obtain the power spectrum for the scalar perturbations. Finally, thus, the power spectrum of scalar perturbations can be parameterized as
\begin{equation}
P_\mathcal{R}(k)=A_\mathcal{R}\left(\frac{k}{k_{pivot}}\right)^{\frac{4(\lambda-3)}{3\lambda-2}}\ ,
\end{equation}
where the amplitude is given by $A_\mathcal{R}=|A|^2A_\Phi$ and $k_{pivot}$ denotes the pivot scale. The scale-invariant power spectrum is given by $\lambda=3$.
This result is different from that one $\lambda=2$ in others' works \cite{Amelino-Camelia:2013wha,Amelino-Camelia:2013tla,Mukohyama:2009gg,Magueijo:2008yk}. Actually, this issue can be demonstrated by the terms proportional to $\frac{d\ln c}{d\ln a}$ in Eq.~(\ref{phi equation}). These terms show that the effective speed of light can be decreased with the expansion of the universe. They lead to the equation of motion (\ref{uk1}) for the scalar perturbations. In this equation, the term proportional to $u_{\bf{k}}$ is different from the one in others' works. Thus, we get different results.

\begin{figure}[!htb]
\includegraphics[width=11 cm]{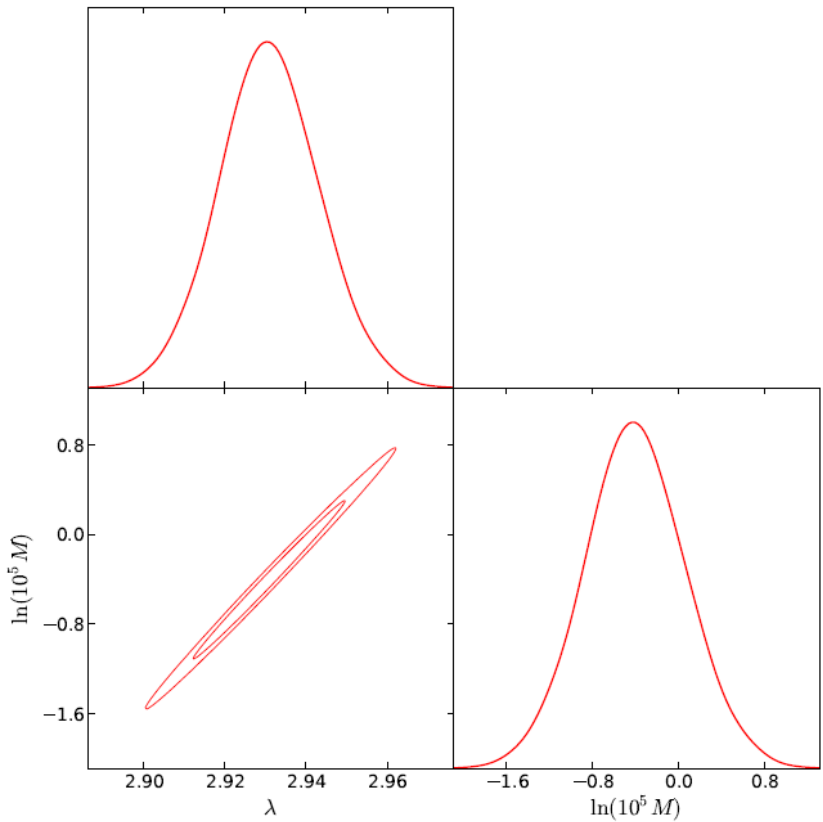}
\caption{The marginalized contour plot and the likelihood distributions of the rainbow index $\lambda$ and energy scale $\ln(10^5M)$ in the gravity's rainbow scenario.}
\label{fig:rainbow}
\end{figure}

The astronomical observations can give certain constraints on the parameters of the gravity's rainbow scenario. In this paper, we use the Planck TT \cite{Ade:2013zuv} and WMAP polarization \cite{Hinshaw:2012aka} datasets to make constraints on the rainbow index $\lambda$ and energy scale $M$ for the gravity's rainbow effects via the CosmoMC \cite{Lewis:2002ah}.
The constraints on $\lambda$ and $M$ are given by
\begin{eqnarray}
\lambda=2.931\pm0.012\ ,~~\ln (10^5M)= -0.401^{+0.457}_{-0.451}\ ,
\end{eqnarray}
at the $68\%$ C.L., respectively. Here the pivot scale is chosen as $k_{\rm{pivot}}=0.05 \rm{Mpc}^{-1}$. Thus, the gravity's rainbow effects would become significant above the energy scale $\sim10^{14}\rm{GeV}$.
In addition, the marginalized contour plot and the likelihood distributions of $\lambda$ and $\ln(10^5M)$ are illustrated in Fig.~\ref{fig:rainbow}.

\section{Conclusions and discussions}

In this paper, we have proposed that the gravity's rainbow scenario could be an alternative of the inflation paradigm of the very early universe. The rainbow function in the metric induces the effective speed of light which depends on the energy of moving particles. We studied the thermodynamics of the system of ultra-relativistic particles with the modified dispersion relation induced by the quantum gravity effects. Then the evolution of the very early universe is determined by the modified Friedmann equation, of which the solution was resolved. Furthermore, we have studied the cosmological linear perturbations and their quantization. The equations for the cosmological perturbations have been derived and the issue of gauge choices was discussed. In the longitudinal gauge, we studied the quantum cosmological perturbations, and then obtained the power spectrum for the primordial comoving curvature perturbations. Furthermore, we make constraints on the rainbow index $\lambda$ and energy scale $M$ of the gravity's rainbow scenario by jointly analyzing the Planck TT and WMAP polarization datasets. Note that the nearly scale-invariant power spectrum for the scalar perturbations required $\lambda\simeq 3$, which satisfies the condition $2/3<\lambda<4$ to account for the flatness and horizon problems.


Though it shed light on the study of the very early universe, our phenomenological scenario still suffers certain puzzling issues. First, the large quantity for the rainbow function is related with a very high energy scale at the start time of the rainbow universe. At such a high energy scale, the quantum gravity effects are unclear. Second, the Einstein's equation should be modified to account for the quantum gravity effects. However, we used the modified Einstein's equation with the speed of light replaced by the effective speed of light. Even though this equation could be reduced back to the conventional one in the general relativity, it should be demonstrated by a consistent theory of quantum gravity in principle. Third, the rainbow metric belongs to the Riemann-Finsler geometry \cite{Book by Mo}, whose dynamics has not been clearly studied so far. In conclusion, one still requires a complete and consistent theory of quantum gravity to study the very early universe in future. Even though there were problems for the gravity's rainbow scenario, our studies still show some interesting results for the research of the very early universe.


\vspace{0.5 cm}

\noindent{\large \bf Acknowledgments}

We acknowledge the use of Planck Legacy Archive, ITP and Lenovo Shenteng 7000 supercomputer in the Supercomputing Center of Chinese Academy of Science for providing computing resources. We are grateful to Prof. Qing-Guo Huang, Xin Li and Yi Ling for useful discussions. The author (S.W.) thanks Dr. Tian-Fu Fu, Yue Huang and Yu-Hang Xing for discussing some details in this paper. He also thanks for the hospitality at the Institute of Astronomy and Space Science in the Sun Yat-Sen University. The author (Z.C.) is funded by the Natural Science Fund of China (NSFC) under Grant No. 11375203, and the author (S.W.) is supported by the project of Knowledge Innovation Program of Chinese Academy of Sciences and grants from NSFC (Grant NO. 11322545 and 11335012).





\begin{thebibliography}{99}
\bibitem{footnote}
  The power spectrum for the statistical thermal fluctuations is proportional to $T^{\frac{3(2-3\lambda)}{2(1+\lambda)}}$ in the gravity's rainbow scenario. When $\lambda\simeq 3$, it will decrease rapidly with the increase of the background temperature. Thus, the thermal fluctuations would be suppressed in this scenario.
\end{thebibliography}
\end{document}